# On the use of light polarization to investigate the size, shape and refractive index dependence of backscattering Ångström exponents

ALAIN MIFFRE*, DANAËL CHOLLETON, PATRICK RAIROUX

*University of Lyon, Université Claude Bernard Lyon 1, CNRS, Institut Lumière Matière, F-69622, Villeurbanne, France*

In this letter, we exploit the polarization property of light to investigate the Ångström exponent describing the wavelength dependence of optical backscatter between two wavelengths. Where previous interpretation of Ångström exponent was that of a particles size indicator, the use of light polarization makes it possible to investigate the Ångström exponent dependence on the particles shape by separately retrieving the backscattering Ångström exponent of the spherical (s) and non-spherical (ns) particles contained in an atmospheric particle mixture (p) = {s, ns}. As an output, analytical solutions of the Maxwell's equations (Lorenz-Mie theory, spheroidal model) can then be applied to investigate the Ångström exponent dependence on the particles size and complex refractive index for each assigned shape. Interestingly, lidar-retrieved vertical profiles of backscattering Ångström exponents specific to s and ns-particles can be used by the optical community to evaluate a range of involved particles sizes and complex refractive indices for both particles shapes, spherical and non-spherical, as we remotely demonstrate on a case study dedicated to a dust nucleation event.

___________________________________________________________________________________

In 1929, while studying the ability of light scattering to reveal the size of atmospheric dust particles, Ångström published a paper [1] where he introduced a quantity that he called $a$ and noted that "the larger the particles, the smaller was the value found for $a$". Nowadays, the Ångström exponent for particles backscattering $BAE_p$, i.e. the value of $a$ in the backward scattering direction, is considered as a qualitative particle size indicator and remotely evaluated every day from atmospheric multi-wavelength lidar instruments [2–4]. By taking benefit from the polarization property of light, this letter aims at developing a new formalism (the heart of which is Eq. 6) allowing to separately retrieve the backscattering Ångström exponent of the spherical (s) and non-spherical (ns) fraction of ambient aerosols contained in a particle mixture (p) = {s, ns}. Interestingly, the lidar-retrieved vertical profiles of these two quantities, retrieved from a 2λ-polarization lidar; allow to evaluate a range of involved particle sizes and complex refractive indices, for s- and ns-particles separately. Basically, the backscattering Ångström exponent describes the wavelength dependence of the particles backscattering coefficient $\beta_p$ between wavelength $\lambda_1$ and wavelength $\lambda_2 > \lambda_1$ [2]:

$$BAE_p = -Ln(\beta_{p,2}/\beta_{p,1})/\Lambda \quad (1)$$

where $\beta_{p,1} = \beta_p(\lambda_1)$, $\beta_{p,2} = \beta_p(\lambda_2)$ and $\Lambda = Ln(\lambda_2/\lambda_1) > 0$. Following Eq. (1), a negative (resp. positive) Ångström exponent means that light backscattering is more (resp. less) important at wavelength $\lambda_2$ compared with wavelength $\lambda_1$. According to the light scattering theory [5], the backscattering Ångström exponent depends on the particles size, shape and $CRI$ as $\beta_p$ does:

$$\beta_{p,i} = \int_{SD} C_{back,p}(r_p, m_p, \lambda_i) n_p(r_p) dr_p \quad (2)$$

where $i = \{1,2\}$, $C_{back,p}$ is the particles backscattering cross-section, $m_p$ is the particles $CRI$, responsible for light absorption, and the integral is performed over the particles size distribution ($SD$), characterized by number density $n_p$ at radius $r_p$. Recent numerical simulations [6] studied the Ångström exponent dependence on the imaginary part of the complex refractive index ($CRI$). Likewise, we computed it for several dust shape models (spheroids, smooth and rough stereograms) assigned to diverse particles mineralogy (Fig. 5 in [7]). However, following Eqs. (1,2), the Ångström exponent dependence on the particles size, shape and $CRI$ is implicit and hence difficult to consider. This is especially true after long-range transport, where particles presenting different sizes, shapes and refractive indices coexist [7,8]. As both spherical ($s$) and nonspherical ($ns$) particles then cohabit, the backscattering Ångström exponent cannot be evaluated from the Lorenz-Mie theory, nor from the spheroidal model whose aspect ratio distribution is optimized for $ns$-particles only [9]. As a work hypothesis, we assume the particles distributions of size, shape and refractive index are statistically independent and can be computed separately [5,6]. Then, instead of working on $BAE_p$, we consider the backscattering Ångström exponent dedicated to $s$ and $ns$-particles, defined by following Eq. (1):

$$BAE_s = -Ln(\beta_{s,2}/\beta_{s,1})/\Lambda \quad (3)$$

$$BAE_{ns} = -Ln(\beta_{ns,2}/\beta_{ns,1})/\Lambda \quad (4)$$

Interestingly, the polarization property of light can be exploited to retrieve the Ångström exponents $BAE_s$ and $BAE_{ns}$ from Eqs. (3, 4) as the $\{s, ns\}$-particles backscattering coefficients $\beta_s$ and $\beta_{ns}$ can be retrieved at wavelength $\lambda_i$ from $\beta_p(\lambda_i)$ as we extensively described in [7,8,10]. Moreover, from Eq. (1), we note that $BAE_p$ is intensive, in contrary to $\beta_p$. Hence, the backscattering Ångström exponent $BAE_p$ of a particles mixture (p) = {s, ns} a priori differs from $BAE_s$ and $BAE_{ns}$, the s and ns-particles backscattering Ångström exponents. The triple subscripts (p, s, ns) are hence important as intrinsic to our new methodology. To establish the relationship between $BAE_p$, $BAE_s$ and $BAE_{ns}$, we first introduce the ratio $X_{ns} = \beta_{ns}/\beta_p$ of ns to p-particles backscattering coefficients. $X_{ns}$ relates to the depolarization ratio $\delta_p$ of the particles mixture $(p) = \{s, ns\}$ and to that $\delta_{ns}$ of its ns-particles [11] as follows [7,8]:

$$X_{ns} = \beta_{ns}/\beta_p = \delta_p(1 + 1/\delta_{ns}) \quad (5)$$

We then consider the ratio $\beta_{p,2}/\beta_{p,1}$ and replace $\beta_{p,2}$ with $\beta_{s,2} + \beta_{ns,2}$. After applying Eqs. (1, 3, 4, 5), we finally get:

$$e^{-\Lambda BAE_p} = e^{-\Lambda BAE_s} + X_{ns,1} \times (e^{-\Lambda BAE_{ns}} - e^{-\Lambda BAE_s}) \quad (6)$$

where $X_{ns,1}$ is the value of $X_{ns}$ at wavelength $\lambda_1$. Alternately, Eq. (6) can be rephrased with $X_{ns,2}$ as follows $e^{\Lambda BAE_p} = e^{\Lambda BAE_s} + X_{ns,2} \times (e^{\Lambda BAE_{ns}} - e^{\Lambda BAE_s})$. Eq. (6) states that $BAE_p$ equals $BAE_s$ (resp. $BAE_{ns}$) only when $X_{ns,1} = 0$ (resp. 1). In the most general case, at a given altitude, $BAE_p$ must be discerned from $BAE_s$ and from $BAE_{ns}$: $BAE_p$ is not dedicated to s or ns-particles but $BAE_s$ and $BAE_{ns}$ are. As a result, $BAE_s$ (resp. $BAE_{ns}$) appears as an optical tracer revealing the wavelength dependence specific to s (resp. $ns$)-particles backscattering at wavelength pair $(\lambda_1, \lambda_2)$. In short, to get $BAE_s$, the light polarization is used as a discriminator, allowing to remove the contribution of $ns$-particles from $BAE_p$, whatever their $SD$ or $CRI$. $BAE_s$ (resp. $BAE_{ns}$) then represents an unequivocal optical signature of the $SD$ and $CRI$ of considered s-particles (resp. $ns$-particles), in contrary to $BAE_p$, at least if polarization and wavelength cross-talks are negligible. By accounting for light polarization, the Ångström exponent dependence on the particles $SD$ and $CRI$ can be quantitatively evaluated for each assigned shape model {s, ns} by means of analytical solutions of the Maxwell's equations: the Lorenz-Mie theory for s-particles backscattering, the spheroidal model [12] for $ns$-particles backscattering. To our knowledge, previous studies [6,13,14] were based on $BAE_p$, which can substantially differ from $BAE_s$ and $BAE_{ns}$, depending on $X_{ns,1}$. Furthermore, s-particles rigorously follow the Lorenz-Mie theory, in contrary to $p$-particles. Therefore, Fig. 1 displays the computed variations of $BAE_s$ (resp. $BAE_{ns}$) at wavelength pair ($\lambda_1$= 355 nm, $\lambda_2$ = 532 nm) as a function of the particles effective radius $r_{eff,s}$ (resp. $r_{eff,ns}$), chosen to characterize the particles $SD$ [15], and the real and imaginary parts of the $CRI$ $m_s$ (resp. $m_{ns}$) of s (resp. $ns$)-particles. In all graphs (including Fig. 2), a mono-modal lognormal $SD$ is considered for s and for $ns$-particles, with geometrical standard deviation $\sigma$ = 1.5 [16], leading to a bimodal $SD$ for $p$-particles. As a first step, refractive indices are assumed wavelength independent and to put light on more striking areas, only Ångström exponents for effective radii above 10 nm are displayed (below, Rayleigh scattering leads to 4).

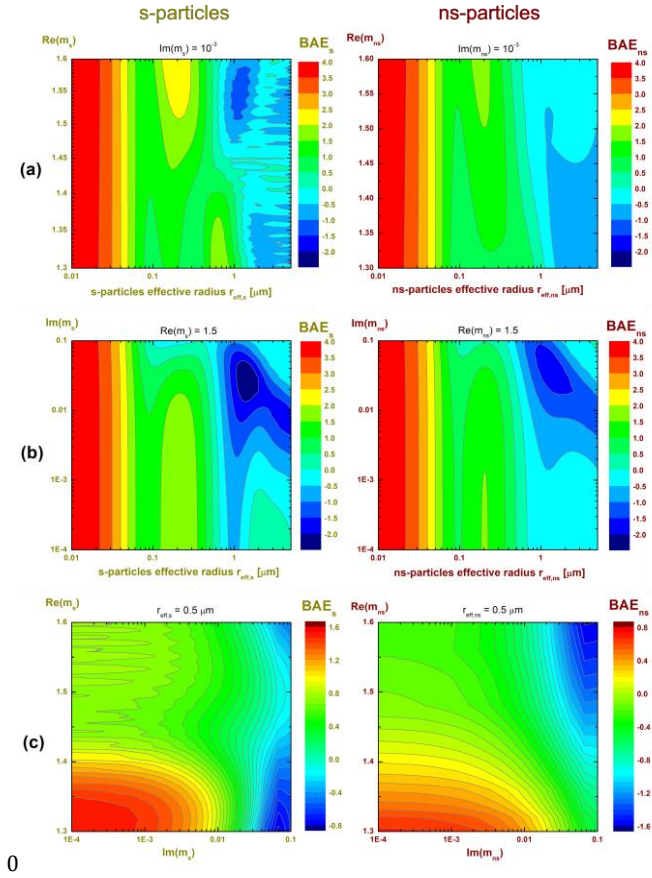

Fig. 1: Contour-plots at constant $BAE_s$ (left) and $BAE_{ns}$ (right) revealing the dependence of backscattering Ångström exponents with the $SD$ (i.e. effective radius) and $CRI$ (i.e. $Re$, $Im$), when neglecting the wavelength dependence of the $CRI$. $BAE_s$ and $BAE_{ns}$ are computed from Eqs (3, 4) by applying the Lorenz-Mie theory for $s$-particles and, for $ns$-particles, the code by Gasteiger and Wiegner [17], based on the T-matrix method [12] and improved geometrical optics [18], with the [19] size independent shape distribution of spheroids. The Eq. (2) size integral is calculated from 0.1 nm to 30 μm. To put light on more striking regions, the color code may differ from one graph to another.

Under these assumptions, Fig. 1a,b (resp. 1c) highlights that the backscattering Ångström exponents $BAE_s$ and $BAE_{ns}$ can be considered as a size indicator (resp. a refractometer) for $s$ and $ns$–particles but care should then be taken as several effective radii (resp. refractive indices) can lead to the same $BAE_s$ and $BAE_{ns}$-values. The consequence of these variations are to be considered on the measured lidar coefficients, after size integration, as developed below. Fig. 1 can then be used to highlight ranges of possible particles effective radii or / and real and imaginary part of $CRI$ corresponding to given values of $BAE_s$ and $BAE_{ns}$. When increasing the $s$ (or $ns$)-particles effective radius, the sign of $BAE_s$ (or $BAE_{ns}$) can change if absorption is sufficient (see the dark blue region in Fig. 1b), thus reflecting the non-monotonous relationship that exists between $BAE_s$ (or $BAE_{ns}$) and $r_{eff,s}$ (or $r_{eff,ns}$). Fig 1b,c straight lines show that absorption can be neglected below $10^{-3}$ imaginary part of $CRI$. However, to draw such a conclusion, the wavelength dependence of the $CRI$ should be accounted for. This is especially true for mineral dust at $\lambda_1$ = 355 nm wavelength and Under these assumptions, Fig. 1a,b (resp. 1c) highlights that the backscattering Ångström exponents $BAE_s$ and $BAE_{ns}$ can be considered as a size indicator (resp. a refractometer) for $s$ and $ns$–particles but care should then be taken as several effective radii (resp. refractive indices) can lead to the same $BAE_s$ and $BAE_{ns}$-values. The consequence of these variations are to be considered on the measured lidar coefficients, after size integration, as developed below. Fig. 1 can then be used to highlight ranges of possible particles effective radii or / and real and imaginary part of $CRI$ corresponding to given values of $BAE_s$ and $BAE_{ns}$. When increasing the $s$ (or $ns$)-particles effective radius, the sign of $BAE_s$ (or $BAE_{ns}$) can change if absorption is sufficient (see the dark blue region in Fig. 1b), thus reflecting the non-monotonous relationship that exists between $BAE_s$ (or $BAE_{ns}$) and $r_{eff,s}$ (or $r_{eff,ns}$). Fig 1b,c straight lines show that absorption can be neglected below $10^{-3}$ imaginary part of $CRI$. However, to draw such a conclusion, the wavelength dependence of the $CRI$ should be accounted for. This is especially true for mineral dust at $\lambda_1$= 355 nm wavelength and smoke at $\lambda_2$ = 532 nm wavelength [6,9,20]. Fig. 2 hence displays the variations of $BAE_s$ (resp. $BAE_{ns}$) as a function of $r_{eff,s}$ (resp. $r_{eff,ns}$) by accounting for the wavelength dependence of $m_s$ and $m_{ns}$ at wavelength pair $(\lambda_1, \lambda_2)$ for several case studies. For $s$-particles, sulfuric acid [21], ammonium sulfate and salt particles at RH = 0.8 (i.e. above their deliquescence point) [22] and an internal mixture of silica (90 % core, [11]) and water (10 % shell, [23]) are considered while silica [11], aluminum oxide, iron oxide (hematite) [24] and natural test dust (i.e. a mixture of the above oxides, [11]) are considered as $ns$-particles proxies. As expected from Rayleigh scattering, at low $r_{eff,s}$, $BAE_s$ is close to 4 and depends on

$F(m_s) = |(m_s{}^2 - 1)/(m_s{}^2 + 2)|^2$ at $(\lambda_1, \lambda_2)$ [25]. When Fig. 2 case studies are involved, Ångström exponents $BAE_s$ (resp. $BAE_{ns}$) above 2.4 can be related to $r_{eff,s}$ (resp. $r_{eff,ns}$). Below 2.4, the $CRI$ must also be taken into account. Negative $BAE_s$-values are obtained for $r_{eff,s}$ above 1 µm. $BAE_{ns}$-values below $-0.5$ are related to either large $r_{eff,ns}$ (above 1 µm) if natural dust is concerned or to lower effective radii if more light absorbing particles are involved, such as hematite (large $Im$ at $\lambda_1$ = 355 nm).

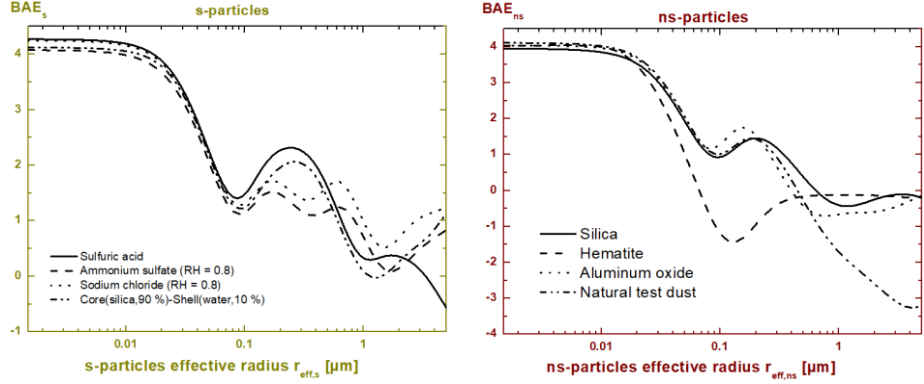

Fig. 2: Computed variations of $BAE_s$ (left) and $BAE_{ns}$ (right) at ($\lambda_1$ = 355 nm, $\lambda_2$ = 532 nm) as a function of $r_{eff,s}$ and $r_{eff,ns}$ for concrete case studies to account for the wavelength dependence of $m_s$ and $m_{ns}$ at wavelength $(\lambda_1, \lambda_2)$ using [21] for sulfuric acid, [22] for ammonium sulfate and salt particles at RH = 0.8, [11] for silica, [24] for hematite and aluminum oxides, [23] for water and [11] for natural test dust.

The real situation is more diverse than Fig. 1 and 2 numerical simulations, so that sensitive and accurate 2λ-polarization lidar atmospheric measurements are required. As a case study, we provide a height-dependent retrieval of $BAE_p$, $BAE_s$ and $BAE_{ns}$ at $(\lambda_1 = 355$ nm, $\lambda_2 = 532$ nm) during a Saharan dust outbreak at Lyon (France) using our dual-wavelength polarization lidar **[26]**. The retrieved profiles of $(\beta_p, \beta_s, \beta_{ns})$ at wavelength pair $(\lambda_1, \lambda_2)$ and corresponding Ångström exponents $(BAE_p, BAE_s, BAE_{ns})$ for p, s, ns-particles backscattering are displayed in Fig. 3a, together with $X_{ns,1}$ and $X_{ns,2}$. Please note that the Ångström exponents are unambiguously retrieved from $(\beta_p, \beta_s, \beta_{ns})$ by applying Eqs. (1, 3, 4) since our 2λ-polarization lidar exhibits a high sensitivity and accuracy [26], i.e. fully negligible polarization and wavelengths cross-talks (1:10$^7$ polarization discriminator, OD5 wavelength discriminator, [27]). For completeness, we recall that $(\beta_p, \beta_s, \beta_{ns})$ are retrieved at altitude z and wavelength $\lambda_i$ by applying the methodology described in [7,8], where the co- $(\beta_{p,//})$ and cross- $(\beta_{p,\perp})$ polarized backscattering coefficients (Fig. 3a) providing $\beta_{p,i}$ are obtained from the lidar raw data after accurate calibration [27] by applying the Klett's algorithm, using constant lidar ratios ($\pm$ 10 sr) taken from the literature [28]. At altitude z and wavelength $\lambda_i$, $(\beta_{s,i}, \beta_{ns,i})$ are obtained by following [7,8], using $\delta_{ns}(\lambda_1)$ = 35.0 % and $\delta_{ns}(\lambda_2)$ = 30.5 % [11,29]. The relationship between $BAE_p$, $BAE_s$ and $BAE_{ns}$ established in Eq. (6) can be verified in our Fig. 3a case study: at every altitude above 2.5 km where the dust cloud is present [27], BAE$_p$ lies in between $BAE_s$ and $BAE_{ns}$ and approaches BAE$_s$ only when $X_{ns,1}$, as provided by Eq. (5), is close to zero. Moreover, over considered altitudes (2.5-4 km), our experimental points ($e^{-\Lambda BAE_p}, X_{ns,1}$) can be adjusted, within our error bars, with a straight line (R$^2$ = 0.94, see Fig. 3b). Interestingly, the slope and the intercept then provide column-mean Ångström exponents BAE$_s$ = 2.3 $\pm$ 0.6 and BAE$_{ns}$ = $-0.8 \pm 0.1$ which are in agreement with the Fig. 3b profiles of BAE$_s$ and BAE$_{ns}$, on average over considered altitudes.

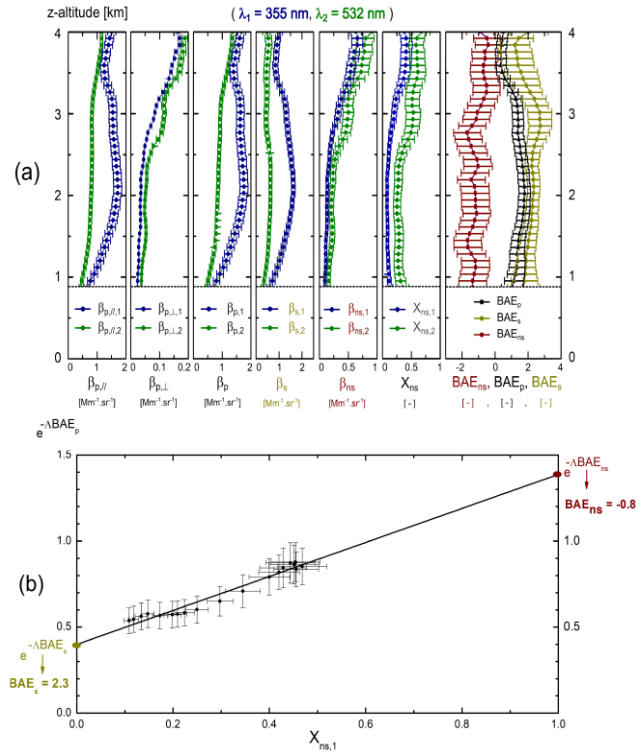

Fig. 3: (a) Lidar profiles of particles backscattering ($\beta_p, \beta_s, \beta_{ns}$) at wavelength pair ($\lambda_1$ = 355 nm in blue, $\lambda_2$ = 532 nm in green) and corresponding Ångström exponents ($BAE_p, BAE_s, BAE_{ns}$) for p, s, ns-particles backscattering during a dust nucleation event at Lyon (France, July 2nd 2015, 14h UTC) taken as a case study. The difference in the ($BAE_p, BAE_s, BAE_{ns}$) profiles originates from $X_{ns,1}$ (or $X_{ns,2}$). (b) $e^{-\Lambda BAE_p}$ as a function of $X_{ns,1}$ as a validation of Eq. (6).

Fig. 1 can be used to discuss on the range of involved / uninvolved effective radii or / and complex refractive indices. For example, to $BAE_s = 2.7 \pm 0.5$, as evaluated at 2.7 km altitude, correspond the yellow and orange regions in Fig. 1a,b which are to be distinguished from bluish areas within our 0.5 error bar. Also, Fig. 1c shows that $r_{eff,s}$ probably differs from 0.5 µm. Likewise, to the value $BAE_{ns} = -1.4 \pm 0.8$ observed at 2.7 km altitude, correspond bluish areas, linked to large effective radius (1 µm) or / and large imaginary part of $m_{ns}$, which are to be distinguished from orange, yellow and green areas within our 0.8 error bar. When the $CRI$ of $s$ (resp. $ns$)-particles is known, Fig. 2 can be applied to highlight a range of effective radii for $s$ and $ns$-particles backscattering by including the wavelength dependence of the $CRI$. In our case study, $s$-particles mainly correspond to sulfuric acid, originating from the subsequent particle growth following a nucleation event promoted by mineral dust [27,30]: fine $s$-particles are then formed in the presence of a few quantity of coarser dust $ns$-particles (hence, $X_{ns,1}$ remains below 0.5 in Fig. 3). The range of retrieved effective radii interestingly reflects this underlying microphysics [27,30,31]: from $BAE_s = 2.7 \pm 0.5$ at 2.7 km altitude, we get $r_{eff,s}$ in the range of either $40 \pm 10$ nm or $250 \pm 80$ nm by considering Fig. 2 for s–particles sulfuric acid. In the meantime, $BAE_{ns} = -1.4 \pm 0.8$ corresponds to larger $r_{eff,ns}$, in the range of $1.1 \pm 0.5$ µm when $ns$–particles natural test dust are considered in Fig. 2. Please note that if $BAE_p$ had been considered instead of $BAE_s$ and $BAE_{ns}$, the range of retrieved effective radii would have been noticeably different: $BAE_p = 1.4 \pm 0.3$ leads to $550 \pm 90$ nm for $r_{eff,s}$ and $200 \pm 50$ nm for $r_{eff,ns}$. Symmetrically, when the particles effective radius is known [6,9], Fig. 1a, b can be used to discuss on the range of involved / uninvolved complex refractive indices. Care should however be taken as Fig. 1 does not account for the wavelength dependence of the $CRI$. When the SD and the $CRI$ of $s$ and $ns$-particles are unknown, a range of possibly involved effective radii and $CRI$ can still be derived from Fig. 1 if differently colored regions can be exhibited. According to Fig. 1, $BAE_s$ and $BAE_{ns}$ should then be experimentally evaluated with an accuracy of 0.5 (at most out of 4). Also, to further disentangle the SD from the $CRI$, our combined use of polarization and spectroscopy may be further applied at a different wavelength pair ($\lambda_1$ = 532 nm, $\lambda_2$ = 1064 nm) or at several wavelengths pairs simultaneously, using 3λ-(or more)-polarization lidar instruments [9,29]. Though our methodology can provide absolute ranges for particle effective radii or refractive indices, more precise values can be obtained by solving the inverse problem, which is however complex as underdetermined [32] and far beyond the scope of this letter. Still as is, the direct problem is interesting in itself, as underlined in [14]. Also, the conclusion of [6] stresses the need to develop the forward model by accurately describing polarization properties at laser backscattering. Moreover, our methodology can be applied in complement to existing inversion algorithms [6] to improve the accuracy of the retrieval by reducing the range of involved effective radii or / and $CRI$ to be considered for $s$ and for $ns$-particles. As a conclusion, use of light polarization provides a deeper understanding of the Ångström exponent of optical backscatter by investigating its dependence on the particles size, real and imaginary part of the $CRI$, for two shapes, corresponding to $s$ and $ns$-particles. If need be, this methodology can be extended to other light scattering numerical models for $ns$-particles backscattering [7] and even to other light scattering directions.

Acknowledgment. We thank T. Mehri for his help in the 2λ-polarization lidar and CNRS and Lyon University for funding.

Disclosures. The authors declare no conflicts of interest.

References
A. Ångström, "On the Atmospheric Transmission of Sun Radiation and on Dust in the Air," Geogr. Ann. 11(2), 156–166 (1929).
2. M. Haarig, A. Ansmann, H. Baars, C. Jimenez, I. Veselovskii, R. Engelmann, and D. Althausen, "Depolarization and lidar ratios at 355, 532, and 1064 nm and microphysical properties of aged tropospheric and stratospheric Canadian wildfire smoke," Atmospheric Chem. Phys. 18(16), 11847–11861 (2018).
3. A. Valenzuela, F. J. Olmo, H. Lyamani, M. Antón, G. Titos, A. Cazorla, and L. Alados-Arboledas, "Aerosol scattering and absorption Ångström exponents as indicators of dust and dust-free days over Granada (Spain)," Atmospheric Res. 154, 1–13 (2015).
4. N. Sugimoto and C. H. Lee, "Characteristics of dust aerosols inferred from lidar depolarization measurements at two wavelengths," Appl. Opt. 45(28), 7468–7474 (2006).
5. M. I. Mishchenko, L. D. Travis, and A. A. Lacis, Scattering, Absorption, and Emission of Light by Small Particles (Cambridge University Press, 2002).
6. I. Veselovskii, P. Goloub, T. Podvin, V. Bovchaliuk, Y. Derimian, P. Augustin, M. Fourmentin, D. Tanre, M. Korenskiy, D. N. Whiteman, A. Diallo, T. Ndiaye, A. Kolgotin, and O. Dubovik, "Retrieval of optical and physical properties of African dust from multiwavelength Raman lidar measurements during the SHADOW campaign in Senegal," Atmospheric Chem. Phys. 16(11), 7013–7028 (2016).
7. T. Mehri, O. Kemppinen, G. David, H. Lindqvist, J. Tyynelä, T. Nousiainen, P. Rairoux, and A. Miffre, "Investigating the size, shape and surface roughness dependence of polarization lidars with light-scattering computations on real mineral dust particles: Application to dust particles' external mixtures and dust mass concentration retrievals," Atmospheric Res. 203, 44–61 (2018).
8. G. David, B. Thomas, T. Nousiainen, A. Miffre, and P. Rairoux, "Retrieving simulated volcanic, desert dust and sea-salt particle properties from two/three-component particle mixtures using UV-VIS polarization lidar and T matrix," Atmospheric Chem. Phys. 13(14), 6757–6776 (2013).
9. D. Müller, I. Veselovskii, A. Kolgotin, M. Tesche, A. Ansmann, and O. Dubovik, "Vertical profiles of pure dust and mixed smoke–dust plumes inferred from inversion of multiwavelength Raman/polarization lidar data and comparison to AERONET retrievals and in situ observations," Appl. Opt. 52(14), 3178 (2013).
10. A. Miffre, G. David, B. Thomas, and P. Rairoux, "Atmospheric non-spherical particles optical properties from UV-polarization lidar and scattering matrix," Geophys. Res. Lett. 38(16), L16804 (2011).
11. A. Miffre, T. Mehri, M. Francis, and P. Rairoux, "UV–VIS depolarization from Arizona Test Dust particles at exact backscattering angle," J. Quant. Spectrosc. Radiat. Transf. 169, 79–90 (2016).
12. M. I. Mishchenko and L. D. Travis, "Capabilities and limitations of a current FORTRAN implementation of the T-matrix method for randomly oriented, rotationally symmetric scatterers," J. Quant. Spectrosc. Radiat. Transf. 60(3), 309–324 (1998).
13. I. Veselovskii, D. N. Whiteman, M. Korenskiy, A. Suvorina, A. Kolgotin, A. Lyapustin, Y. Wang, M. Chin, H. Bian, T. L. Kucsera, D. Pérez-Ramírez, and B. Holben, "Characterization of forest fire smoke event near Washington, DC in summer 2013 with multi-wavelength lidar," Atmospheric Chem. Phys. 15(4), 1647–1660 (2015).
14. S. P. Burton, E. Chemyakin, X. Liu, K. Knobelspiesse, S. Stamnes, P. Sawamura, R. H. Moore, C. A. Hostetler, and R. A. Ferrare, "Information content and sensitivity of the 3 β + 2 α lidar measurement system for aerosol microphysical retrievals," Atmospheric Meas. Tech. 9(11), 5555–5574 (2016).
15. J. E. Hansen and L. D. Travis, "Light scattering in planetary atmospheres," Space Sci. Rev. 16(4), 527–610 (1974).
16. G. L. Schuster, O. Dubovik, and B. N. Holben, "Angstrom exponent and bimodal aerosol size distributions," J. Geophys. Res. 111(D7), (2006).
17. J. Gasteiger and M. Wiegner, "MOPSMAP v1.0: a versatile tool for the modeling of aerosol optical properties," Geosci. Model Dev. 11(7), 2739–2762 (2018).
18. P. Yang, Q. Feng, G. Hong, G. W. Kattawar, W. J. Wiscombe, M. I. Mishchenko, O. Dubovik, I. Laszlo, and I. N. Sokolik, "Modeling of the scattering and radiative properties of nonspherical dust-like aerosols," J. Aerosol Sci. 38(10), 995–1014 (2007).
19. K. Kandler, L. SchüTZ, C. Deutscher, M. Ebert, H. Hofmann, S. JäCKEL, R. Jaenicke, P. Knippertz, K. Lieke, A. Massling, A. Petzold, A. Schladitz, B. Weinzierl, A. Wiedensohler, S. Zorn, and S. Weinbruch1, "Size distribution, mass concentration, chemical and mineralogical composition and derived optical parameters of the boundary layer aerosol at Tinfou, Morocco, during SAMUM 2006," Tellus B Chem. Phys. Meteorol. 61(1), 32–50 (2009).
20. A. Ansmann, A. Petzold, K. Kandler, I. Tegen, M. Wendisch, D. Müller, B. Weinzierl, T. Müller, and J. Heintzenberg, "Saharan Mineral Dust Experiments SAMUM–1 and SAMUM–2: what have we learned?," Tellus B Chem. Phys. Meteorol. 63(4), 403–429 (2011).
21. J. R. Hummel, E. P. Shettle, and D. R. Longtin, A New Background Stratospheric Aerosol Model for Use in Atmospheric Radiation Models (OPTIMETRICS INC BURLINGTON MA, 1988).
22. M. I. Cotterell, R. E. Willoughby, B. R. Bzdek, A. J. Orr-Ewing, and J. P. Reid, "A complete parameterisation of the relative humidity and wavelength dependence of the refractive index of hygroscopic inorganic aerosol particles," Atmospheric Chem. Phys. 17(16), 9837–9851 (2017).
23. G. M. Hale and M. R. Querry, "Optical Constants of Water in the 200-nm to 200-μm Wavelength Region," Appl. Opt. 12(3), 555–563 (1973).
24. M. R. Querry, Optical Constants (MISSOURI UNIV-KANSAS CITY, 1985).
25. C. F. Bohren and D. R. Huffman, Absorption and Scattering of Light by Small Particles (Wiley-VCH, 1983).
26. G. David, A. Miffre, B. Thomas, and P. Rairoux, "Sensitive and accurate dual-wavelength UV-VIS polarization detector for optical remote sensing of tropospheric aerosols," Appl. Phys. B 108(1), 197–216 (2012).
27. A. Miffre, D. Cholleton, T. Mehri, and P. Rairoux, "Remote Sensing Observation of New Particle Formation Events with a (UV, VIS) Polarization Lidar," Remote Sens. 11(15), 1761 (2019).
28. I. Veselovskii, O. Dubovik, A. Kolgotin, T. Lapyonok, P. Di Girolamo, D. Summa, D. N. Whiteman, M. Mishchenko, and D. Tanré, "Application of randomly oriented spheroids for retrieval of dust particle parameters from multiwavelength lidar measurements," J. Geophys. Res. 115(D21), (2010).
29. M. Tesche, A. Kolgotin, M. Haarig, S. P. Burton, R. A. Ferrare, C. A. Hostetler, and D. Mueller, "3+2 + X: what is the most useful depolarization input for retrieving microphysical properties of non-spherical particles from lidar measurements using the spheroid model of Dubovik et al. (2006)?," (2019).
30. Y. Dupart, S. M. King, B. Nekat, A. Nowak, A. Wiedensohler, H. Herrmann, G. David, B. Thomas, A. Miffre, P. Rairoux, B. D'Anna, and C. George, "Mineral dust photochemistry induces nucleation events in the presence of SO2," Proc. Natl. Acad. Sci. 109(51), 20842–20847 (2012).
31. G. David, B. Thomas, Y. Dupart, B. D'Anna, C. George, A. Miffre, and P. Rairoux, "UV polarization lidar for remote sensing new particles formation in the atmosphere," Opt. Express 22(S3), A1009 (2014).

32.	E. Chemyakin, S. Burton, A. Kolgotin, D. Müller, C. Hostetler, and R. Ferrare, "Retrieval of aerosol parameters from multiwavelength lidar: investigation of the underlying inverse mathematical problem," Appl. Opt. 55(9), 2188–2202 (2016).